# Vibrational properties of photochromic yttrium oxyhydride and oxydeuteride thin films

M. Zubkins[1*], J. Gabrusenoks[1], R. Aleksis[2], G. Chikvaidze[1], E. Strods[1], V. Vibornijs[1], A. Lends[2], K. Kundzins[1], J. Purans[1]

[1]Institute of Solid State Physics, University of Latvia, Kengaraga 8, LV-1063, Riga, Latvia

[2]Latvian Institute of Organic Synthesis, Aizkraukles 21, LV-1006, Riga, Latvia

*Corresponding author: martins.zubkins@cfi.lu.lv

**Abstract**

A comprehensive study of the vibrational properties of photochromic yttrium oxyhydride (YHO) and oxydeuteride (YDO) thin films is presented. These films are deposited using reactive magnetron sputtering, followed by post-oxidation. Our investigation employs vibrational Fourier-transform infrared (FTIR) spectroscopy, in conjunction with first-principles Density Functional Theory (DFT) calculations. The FTIR spectra of the films reveal broad vibrational bands, primarily attributed to the disordered structure containing small crystallites (<10 nm), as confirmed by solid-state nuclear magnetic resonance and X-ray diffraction measurements. An isotopic shift from approximately 900 to 745 cm$^{-1}$ is observed in the hydrogen/deuterium-related vibration band, while the lower frequency bands (< 600 cm$^{-1}$) remain unaffected upon replacement of hydrogen with deuterium. These experimental observations are consistent with the DFT theoretical calculations for various stable YHO lattices reported in the literature. Illumination of the films with ultraviolet light at 3.3 eV leads to additional absorption not only in the visible light range but also up to approximately 2000 cm$^{-1}$ in the mid-infrared region. However, no phase transformation change or formation of hydroxyl (OH) groups are observed following illumination. Our findings provide valuable insight into the vibrational and photochromic properties of YH(D)O thin films.

**Introduction**

Photochromic materials, capable of undergoing reversible color changes upon exposure to light with specific energy, have garnered significant attention for their potential applications in various fields, including smart windows, optical devices, and information storage [1,2]. Among these materials, oxygen-containing yttrium hydride, often referred to as yttrium oxyhydride (YHO), thin films have emerged as promising candidates, owing to their intriguing color-neutral photochromic properties [3–5]. YHO is a mixed-anion compound [6] with the coexistence of both $O^{2-}$ and $H^-$ ions. Experimental studies have confirmed the formation of a face-centered cubic cation lattice [7–9]. YHO films become opaque when exposed to UV-visible light and return to a transparent state in a couple of minutes to several days, depending on illumination conditions. The relatively long bleaching time implies that the underlying process is not purely electronic. The photochromic properties are highly dependent on the O:H concentration ratio [4,10,11], the anionic ordering in the sublattice [12,13], and on defects [14].

Reactive magnetron sputtering, combined with post-oxidation of $YH_{2-x}$ in an air environment, can be used to deposit YHO films. The extent of oxidation depends on the porosity of the structure, a parameter controllable through deposition pressure [5]. As-deposited photochromic YHO films exhibit a yellowish appearance due to increased absorption on the blue part of the visible spectrum, with an optical band gap approximately in the range of 2.5–3.2 eV [15,16]. During the oxidation of $YH_{2-x}$, oxygen occupies free tetrahedral sites, while excess hydrogen can migrate to the more unstable octahedral sites [17], with a portion of it even being released from the compound [18]. Calculation studies have shown that a random arrangement of anions in the tetrahedral sites is believed to be more energetically favorable compared to a periodic arrangement [19]. Furthermore, hydrogen anions can occupy positions at intermediate locations between the ideal tetrahedral and octahedral lattice sites after structural relaxation [19]. The disordered configuration of anions may also be influenced by the synthesis method and oxidation conditions. Anion-disordered structures hinder ionic mobility, limiting ion diffusion to local scales rather than long-range distances [20]. The solid-state nuclear magnetic resonance (ssNMR) spectroscopy studies have revealed a heterogeneous structure in YHO films, with regions exhibiting varying hydrogen content – some rich and some poor in hydrogen – along with several distinct hydrogen environments [12]. The precise mechanism of the photochromic effect, however, remains under investigation. It has been suggested recently that photo-excited electrons reduce trivalent yttrium, forming metallic clusters, while holes are trapped by hydride anions [14]. The kinetics of mobile hydrogen also appear to play a pivotal role in this process [21]. Different types of anionic isotopes ($^{17}O$, $^{18}O$, D) have been employed either to explore long-range diffusion within the film and the exchange of material with the environment [22,23], or to enhance the NMR signal for determining structural information along with anionic dynamics [12]. However, it seems that none of these isotopes significantly affect the photochromic effect.

In this study, we examine the vibrational characteristics of YHO and anion-labeled YDO thin films, exploring their structural properties and the photochromic effect. Vibrational Fourier-transform infrared (FTIR) spectroscopy is suitable for investigating the structure and coordination modes of oxyhydride compounds and is used for the first time in the far-infrared region for YHO films. Deuterium, rather than hydrogen, is utilized to label hydrogen and identify the associated vibration bands. Using a combination of experimental FTIR spectroscopy, ssNMR, and X-ray diffraction (XRD) techniques, along with density functional theory (DFT) calculations, we have gained valuable insights into the behavior of these films.

## 2. Experimental details

The film deposition of YH(D)O was carried out using a custom made vacuum PVD coater G500M from Sidrabe Vacuum, Ltd. The $YH(D)_2$ films were deposited onto Si substrates by reactive pulsed direct current (pulsed-DC) (80 kHz, 2.5 µs reverse time) magnetron sputtering from an Y (purity 99.9%) target in an Ar (purity 99.9999%) and $H_2/D_2$ (purity 99.999%/99.8%) atmosphere, maintaining a $H_2/D_2$ to Ar gas flow ratio of 1/2. A planar balanced magnetron with target dimensions of 150 mm × 75 mm × 3 mm was used. The magnetron was positioned at a distance of 10 cm from the grounded substrate holder, with the substrates facing the target axis. The substrates were not intentionally heated during the deposition process. Before initiating the deposition, the chamber was evacuated to a pressure of $6 \times 10^{-4}$ Pa using a turbo-molecular pump backed by a rotary pump. The target was sputtered at a constant average power of 200 W. Deposition was carried out for 30 minutes at varying sputtering pressures, ranging from 0.75 to 2.65 Pa, by adjusting the pumping speed with a throttle valve. A pressure of 0.75 Pa is close to the critical pressure defined in Ref. [5], below which opaque and non-photochromic films are obtained.

To achieve the YHO phase, oxygen gas (with a purity of 99.999%) was deliberately introduced into the chamber immediately after deposition, reaching approximately 400 Pa over a 30-minute period (dosage ≈ $10^9$ Langmuir) that is enough to obtain YHO phase according to Ref. [15]. Subsequently, most of the YHO samples were capped with an approximately 300 nm thick layer of aluminum within the same vacuum chamber to protect the films from the ambient environment, as similarly performed in Ref. [24]. Characterisation of the thin films using X-ray diffraction (XRD), FTIR, scanning electron microscopy (SEM), and spectroscopic ellipsometry (SE) techniques was conducted within a few days after exposure to air.

FTIR absorbance spectra were measured using a VERTEX 80v vacuum FTIR spectrometer. The experiments were conducted in the range of 30 to either 4000 or 7000 $cm^{-1}$, with the interferometer operating both under vacuum conditions and air, and at a resolution of 4 $cm^{-1}$. Uncoated Si was used as a background reference. The non-capped films were measured in transmittance mode, while the Al-capped films were measured in reflectance mode by irradiating through the substrate. The FTIR spectrum of a darkened YHO film deposited at 0.95 Pa was measured after prolonged illumination for 12 hours with UVA-violet light from a 15 W (≈2.4 $mW/cm^2$) lamp under ambient conditions. The energy of the source light at maximum intensity is 3.3 eV with a spectral width of 0.13 eV.

The crystallographic structure of the films was examined by XRD, using a Rigaku MiniFlex600 X-ray powder diffractometer with Bragg-Brentano θ-2θ geometry and a 600 W Cu anode (Cu Kα radiation, λ = 1.5406 Å) X-ray tube.

The surface and cross-section of the films was visualized using Helios 5 UX dual-beam SEM form Thermo Fisher Scientific. The cross-section images were obtained by scratching the coating off the substrate and imaging scratched pieces on the substrate, which are placed closely perpendicular to the substrate.

Film thicknesses were determined using a SE WOOLLAM RC2 spectroscopic ellipsometer in the spectral range of 550–1690 nm (2.25–0.73 eV), where the films exhibit high transparency. The Cauchy model was applied in this range for the analysis. The primary ellipsometric angles, Ψ and Δ, were measured at incident angles ranging from 55° to 65° in 5° increments. Spectroscopic ellipsometry (SE) experimental data and model-based regression analyses were processed using the WOOLLAM software, CompleteEASE. The mean squared error (MSE) values for the models ranged between 1 and 10. The approximate film thickness in this study was 500 nm. A detailed analysis of the SE measurements over the broader spectral range of 5.9–0.7 eV, employing physical oscillator models, along with the optical constants, their gradients, and the photochromic properties of YHO films deposited under the described conditions, is provided in Ref. [15].

For ssNMR measurements, additional YHO powder samples of approximately 60 mg were synthesised by prolonged deposition for 4 hours on 10 × 10 cm glass substrates to produce a film approximately 4 µm thick. This thickness was chosen not only to enable removal of the film from the substrate but also to ensure sufficient material for obtaining a reliable and high-quality signal from ssNMR. The $^2$H ssNMR spectra were acquired on an 18.8 T Bruker Avance III NMR spectrometer operating at a Larmor frequency of 122.84 MHz at static or 20 kHz magic angle spinning (MAS) regime using a 3.2 mm Bruker H/F X probe at 273 K. The radio-frequency (RF) amplitude of all pulses was 54 kHz. The adiabatic shifting *d*-echo experiment [25] used high-power adiabatic pulses [26,27] with the tanh/tan [28] pulse scheme with 5 MHz sweep width and 50 µs pulse length. The $^2$H shifts were referenced externally to $CH_3OD$ at 3.70 ppm. The data were processed and analysed using the Bruker Topspin 3.4.5 program and in-house written Matlab [29] scripts. Additional experimental ssNMR details can be found in supplementary information (SI).

## 3. First-principles calculations

Calculations of the fundamental vibrations of YHO have been performed using hybrid exchange density functional theory to determine the equilibrium geometries and phonon frequencies. YHO was modelled using linear combination of atomic orbitals (LCAO) methods within the framework of the hybrid density functional approach. To perform hybrid LCAO calculations, we used the periodic CRYSTAL17 code [30], which employs Gaussian-type functions cantered on atomic nuclei as the basis set for the expansion of crystalline orbitals. The full-electron basis sets for O and H and valence-electron-only calculations with the aid of effective core pseudopotentials for Y used in this study were taken from the CRYSTAL basis set library. The B3LYP functional was used in the calculations.

## 4. Results and discussion

*4.1. Structural characterisation – XRD, SEM, ssNMR*

The X-ray patterns of the both Al-capped and non-capped YH(D)O films on Si substrates measured between 20° and 60° of 2θ are shown in Fig. 1. The XRD peaks observed at approximately 29°, 33°, 48°, and 57° correspond to the cubic lattice planes (111), (200), (220), and (311), respectively, indicating the polycrystalline nature of the YH(D)O films. According to the powder diffraction card ICDD 04-002-6939 of $\beta$-YH$_2$ ($Fm$-$3m$, $a$=5.2 Å), the maximums are strongly shifted to lower angles indicating an expansion of the lattice after the incorporation of oxygen, which has already been shown in other studies [24,31–33]. The lattice parameter $a$ ranges between 5.3 and 5.4 Å depending on the deposition conditions without a clear correlation. However, the lattice parameters are close to the values to the stable YHO lattice geometries ($F$-$43m$ and $P$-$43m$) theoretically predicted in Ref. [34]. Additionally, it should be noted that deposition through reactive magnetron sputtering is a non-equilibrium process, and some level of disorder can be anticipated compared to ideal structures. As per the Scherrer equation, the average crystallite size for all samples falls within the range of 5–10 nm.

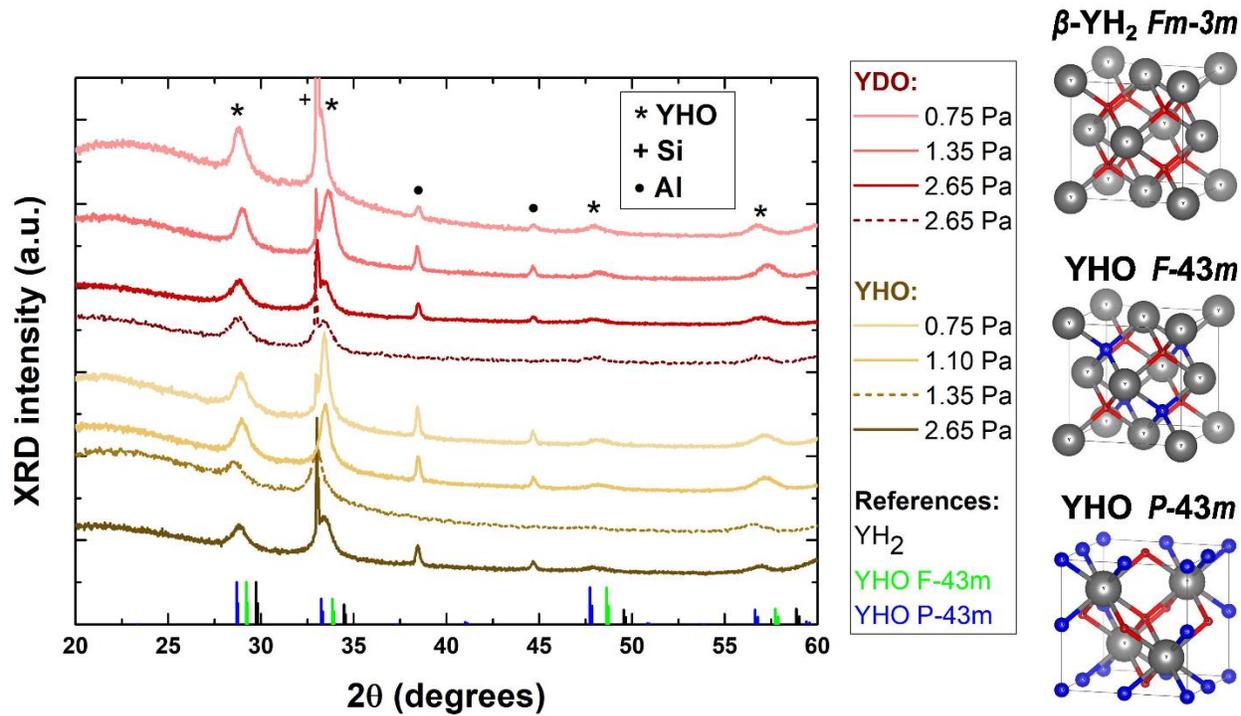

**Fig. 1.** XRD patterns of the both Al-capped (solid lines) and non-capped (dashed lines) YH(D)O films produced by reactive (H$_2$(D$_2$)/Ar gas flow ratio of 1/2) pulsed-DC magnetron sputtering at different deposition pressures. The angles of the Bragg peaks for the cubic $\beta$-YH$_2$ (ICDD 04-002-6939) and the cubic YHO phases from Ref. [34] are shown as references.

Surface morphology and cross-sectional images of the non-capped and Al-capped YHO films deposited at 0.75 Pa, obtained by electron microscopy, are depicted in Fig. 2. The surface of the non-capped film exhibits grainy features with sharp edges and an average size of around 50 nm (Fig. 2(a)), with these grains consisting of smaller crystallites as observed in XRD. The cross-sectional image (Fig. 2(b)) reveals a grained structure of textured and fibrous grains, oriented in the direction of the arriving material flux during growth. This is attributed to low adatom mobility, resulting in continued nucleation of grains. However, the grains gradually enlarge as deposition progresses, forming a V-type shape. This grain formation indicates growth in either zone 1 or zone T (transition), as per the Thornton structure zone diagram [35]. V-shaped grains can form when surface diffusion exceeds grain boundary diffusion, leading to competitive grain growth and a non-homogeneous grain structure throughout the film thickness [36]. The high density of pores at the grain boundaries is evident in the cross-sectional SEM images (Fig. 2(b)), where voids between the fibrous grains are clearly visible. Additionally, the ends of the grains create protrusions on the surface, leading to slight roughness. Fig. 2(c) depicts a dense and homogeneous Al capping layer on top of the YHO film.

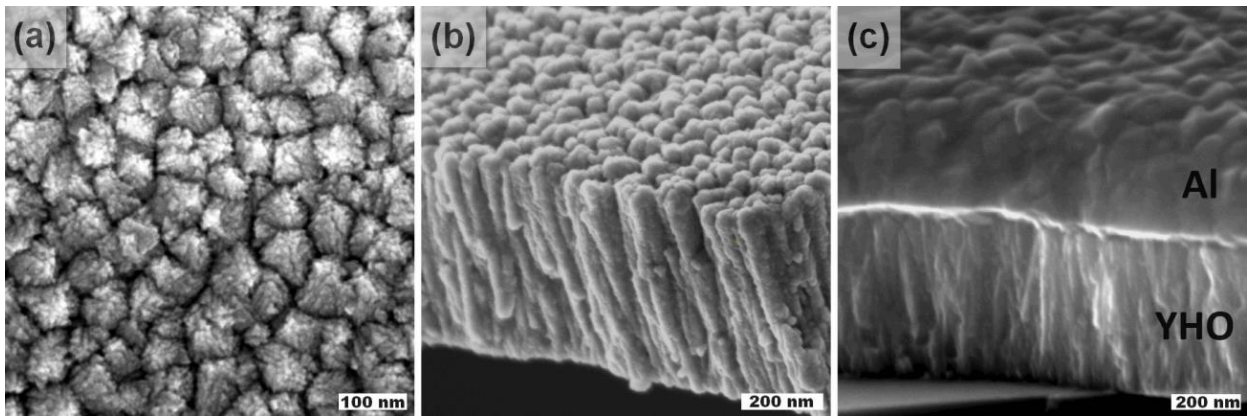

**Fig. 2.** (a) Surface and (b) cross-sectional SEM images of a non-capped YHO film, and (c) the cross-sectional SEM image of an Al-capped YHO film on Si substrates, deposited by reactive ($H_2$/Ar gas flow ratio of 1/2) pulsed-DC magnetron sputtering at a pressure of 0.75 Pa.

We utilised $^2$H detected ssNMR spectroscopy to gain insights on the local hydride structure in YDO. The $^2$H nucleus has a nuclear electric quadrupole moment; hence it is subject to the quadrupolar interaction, which can be used as a highly sensitive probe to monitor changes in the coordination geometry and local structure [37]. We initially examined YDO sample (0.75 Pa) using 1D and 2D static ssNMR methods (see Fig. 3(a,b)). The 1D static $^2$H spectrum shows two distinct $^2$H environments – one with a narrow lineshape (site I in Fig. 3(a)) and a long transverse relaxation constant $T_2$', while the other with a much broader lineshape (site II in Fig. 3(a)) and substantially shorter $T_2$' (see Table S1 in SI). Based on a previous study [12] the narrow component can be assigned to trapped molecular $D_2$ in the crystal structure, while the broader $^2$H site arises from hydride anions. The $^2$H hydride lineshape appears as a Gaussian distribution, which can indicate high degree of structural disorder. Similar featureless $^2$H lineshapes have been observed due to the convolution between the shift and quadrupolar interactions, therefore we carried out the adiabatic shifting *d*-echo experiment, which separates and correlates the two interactions (see Fig. 3(b)) [25]. The 1D projections (Fig. 3(b)) of the experiment show pure shift and quadrupole spectra. We notice that both spectra consist of featureless, Gaussian-like lineshapes, revealing a distribution of $^2$H environments either due to poor crystallinity or anion disorder. Despite the disorder we note that the maximum quadrupole coupling constant $C_Q$ is ca. 15 kHz, which indicates a very symmetric electronic environment as expected for a tetrahedral site in a cubic lattice. Finally, we acquired a 1D experiment under magic-angle spinning (MAS), which averages the anisotropic interactions and concentrates the signal into discrete sidebands, thus increasing the sensitivity and spectral resolution [38]. The 1D MAS spectrum shows an additional very broad $^2$H site spanning ~360 kHz (site III in Fig. 3(c)). Despite the improved resolution under MAS, the sites cannot be resolved based on chemical shifts as the isotropic shift positions of all components are within the linewidth of the centreband; hence we differentiate the sites according to their different anisotropies, i.e., the broadness of each site (see Fig. 3(c)). The spinning sideband pattern of the third $^2$H environment matches well with a random distribution of quadrupole tensor parameters ($C_Q$=140–230 kHz, $\eta_Q$=0.0–1.0), further indicating disorder in the studied samples. The large $C_Q$ values reveal that the electronic environment is of low symmetry. We conjecture that this could be a hydride anion in a severely distorted tetrahedral site or an impurity, such as an $OH^-$ group, which has been observed in YHO previously [12]. However, for an unambiguous assignment a more thorough investigation should be carried out, which is out of the scope of the current study.

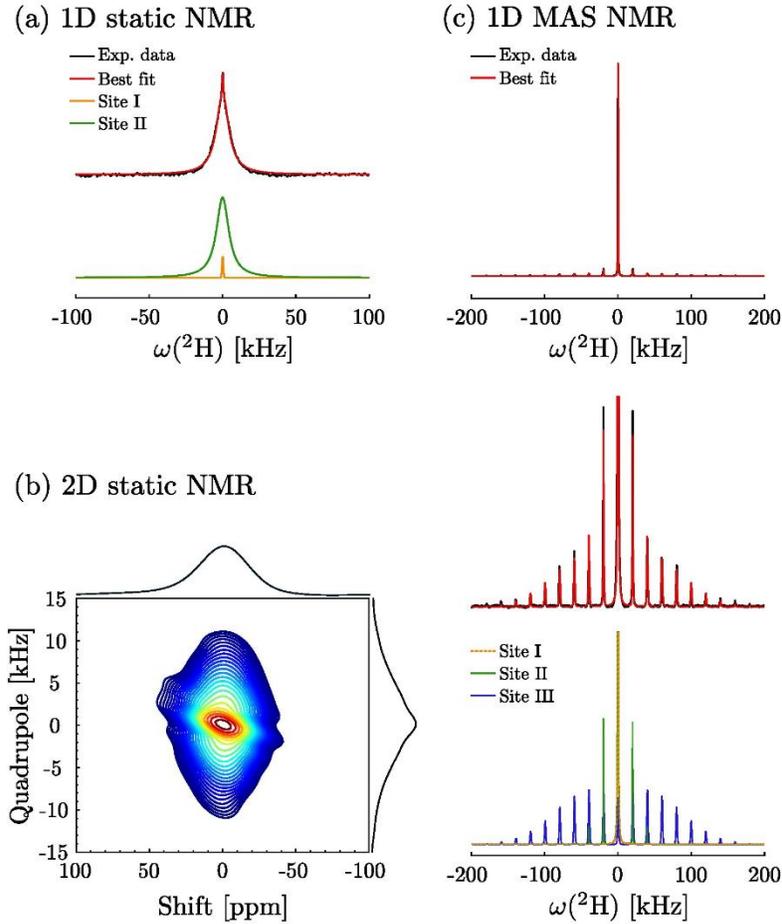

**Fig. 3.** $^2$H ssNMR spectra of YDO (0.75 Pa). (a) 1D static $^2$H experimental and best-fit spectra. (b) Experimental $^2$H 2D adiabatic shifting *d*-echo spectrum. The projections onto the axis show spectra resulting from shift or quadrupolar interactions. (c) 1D 20 kHz MAS experimental and best-fit spectra. The spectra are deconvolved assuming three $^2$H sites with distinct quadrupole anisotropies.

*4.2. First-principles calculations*

Calculations were performed for five different YH(D)O polymorphs with space groups *F-43m*, *Pnma*, *P-43m*, *R-3m*, *P4/nmm*. The calculated vibrations, which can be observed in Raman and IR spectra, correspond to long-wavelength vibrations with a wave vector (q = 0). The structures were selected from Ref. [34] based on the stability of the crystalline phases for the stoichiometric YHO compound. Table 1 contains information about lattice vibration frequencies $\omega_H$ in YHO and their changes when replacing hydrogen with deuterium – $\omega_D$. If the frequency squared ratio $\omega_H^2/\omega_D^2$ is equal to 1, the frequency of this vibration is determined by O and/or Y movement rather than the H(D) sublattice. However, if the ratio equals 2, only the H(D) sublattice participates in the corresponding oscillation. A deviation from the ratio value of 2 indicates the contribution of Y and/or O atoms to such a vibration. As can be seen from Table 1, the vibrations fall into two regions. In the low-frequency region ($\omega \lesssim 800$ cm$^{-1}$), where the ratio $\omega_H^2/\omega_D^2$ is practically equal to 1, and in the high-frequency region ($\omega \gtrsim 800$ cm$^{-1}$), where the ratio is practically equal to 2. Only in the case of the structure *P-43m*, the two oscillations on the boundary of these regions exchange places, i.e. $\omega_H$. So, the phonon spectrum of YH(D)O structures splits into two regions: a high-frequency region dominated by H(D) oscillations, and a low-frequency region dominated by Y and O oscillations. The remaining three columns of Table 1 give the vibrational symmetry and the vibrational activity (A – active, I – inactive) in infrared (IR) and Raman (R) spectra. Information from the CRYSTAL17 program regarding the optimised structures and the lattice vibration frequencies of five YH(D)O phases is summarised in the SI.

**Table 1.** Calculated frequencies of the phonon modes in different YH(D)O polymorphs – *F-43m*, *Pnma*, *P-43m*, *R-3m*, and *P4/nmm*.

| Frequency (cm$^{-1}$) | | $\omega_H^2/\omega_D^2$ | Symmetry | Activity | |
|---|---|---|---|---|---|
| $\omega_H$ | $\omega_D$ | | | IR | Raman |
| *F-43m* (No.216) | | | | | |
| 353.9 | 350.5 | 1.02 | F | A | A |
| 1106.6 | 796.8 | 1.93 | F | A | A |
| *Pnma* (No.62) | | | | | |
| 129.2 | 129.2 | 1.00 | A$_u$ | I | I |
| 135.1 | 135.1 | 1.00 | A$_g$ | I | A |
| 146.1 | 146.1 | 1.00 | B$_{2g}$ | I | A |
| 165.4 | 164.5 | 1.01 | B$_{2u}$ | A | I |
| 172.8 | 172.8 | 1.00 | B$_{1g}$ | I | A |
| 178.8 | 178.3 | 1.01 | $A_g$ | I | A |
| 206.6 | 205.2 | 1.01 | $B_{1u}$ | A | I |
| 210.4 | 210.3 | 1.00 | $B_{3g}$ | I | A |
| 285.0 | 279.4 | 1.04 | $B_{3g}$ | I | A |
| 300.6 | 296.7 | 1.03 | $B_{3g}$ | I | A |
| 316.7 | 312.5 | 1.03 | $B_{2u}$ | A | I |
| 337.6 | 330.5 | 1.04 | $A_g$ | I | A |
| 354.3 | 349.0 | 1.03 | $B_{1u}$ | A | I |
| 375.4 | 372.9 | 1.01 | $B_{3u}$ | A | I |
| 383.3 | 381.9 | 1.01 | $A_u$ | I | I |
| 395.7 | 383.2 | 1.07 | $B_{1u}$ | A | I |
| 426.7 | 417.4 | 1.04 | $A_g$ | I | A |
| 427.8 | 421.2 | 1.03 | $B_{2u}$ | A | I |
| 433.1 | 431.8 | 1.01 | $B_{2g}$ | I | A |
| 439.7 | 439.5 | 1.00 | $B_{1g}$ | I | A |
| 562.0 | 550.5 | 1.04 | $B_{3g}$ | I | A |
| 820.6 | 592.7 | 1.92 | $B_{2u}$ | A | I |
| 880.9 | 645.5 | 1.86 | $A_g$ | I | A |
| 912.2 | 650.0 | 1.97 | $A_u$ | I | I |
| 916.4 | 657.9 | 1.94 | $B_{3u}$ | A | I |
| 925.3 | 683.7 | 1.83 | $B_{1u}$ | A | I |
| 936.7 | 698.5 | 1.80 | $B_{3g}$ | I | A |
| 1124.7 | 803.1 | 1.96 | $A_g$ | I | A |
| 1137.6 | 818.1 | 1.93 | $B_{2u}$ | A | I |
| 1145.4 | 822.5 | 1.94 | $B_{1u}$ | A | I |
| 1151.8 | 825.5 | 1.95 | $B_{1g}$ | I | A |
| 1153.1 | 828.6 | 1.94 | $B_{3g}$ | I | A |
| 1158.9 | 833.1 | 1.94 | $B_{2g}$ | I | A |
| | | *P*-43*m* (No.215) | | | |
| 145.2 | 145.2 | 1.00 | $F_1$ | I | I |
| 170.5 | 170.4 | 1.00 | $F_2$ | A | A |
| 287.5 | 277.8 | 1.07 | E | I | A |
| 294.8 | 287.5 | 1.05 | $F_2$ | A | A |
| 308.2 | 308.2 | 1.00 | A | I | A |
| 331.6 | 328.7 | 1.02 | $F_1$ | I | I |
| 377.9 | 368.1 | 1.05 | $F_2$ | A | A |
| 535.8 | 406.0 | 1.74 | $F_2$ | A | A |
| 611.5 | 600.9 | 1.04 | $F_2$ | A | A |
| 987.7 | 707.1 | 1.95 | $F_2$ | A | A |
| 1064.1 | 762.2 | 1.95 | $F_1$ | I | I |
| 1107.2 | 811.4 | 1.86 | $F_2$ | A | A |
| | | *R*-3*m* (No. 166) | | | |
| 129.3 | 129.3 | 1.00 | $E_g$ | I | A |
| 229.4 | 229.3 | 1.00 | $A_{1g}$ | I | A |
| 355.8 | 351.3 | 1.03 | $A_{2u}$ | A | I |
| 373.4 | 364.8 | 1.05 | $E_u$ | A | I |
| 433.2 | 428.7 | 1.02 | $E_g$ | I | A |
| 457.1 | 456.6 | 1.00 | $A_{1g}$ | I | A |
| 826.4 | 603.2 | 1.88 | $E_u$ | A | I |
| 991.8 | 711.8 | 1.94 | $E_g$ | I | A |
| 1110.2 | 801.9 | 1.92 | $A_{2u}$ | A | I |
| 1148.1 | 816.3 | 1.98 | $A_{1g}$ | I | A |
| *P*4/*nmm* (No.129) | | | | | |
| 160.0 | 159.9 | 1.00 | $E_g$ | I | A |
| 253.2 | 253.2 | 1.00 | $A_{1g}$ | I | A |
| 301.1 | 301.1 | 1.00 | $E_u$ | A | I |
| 318.7 | 316.2 | 1.02 | $A_{2u}$ | A | I |
| 383.9 | 383.7 | 1.00 | $E_g$ | I | A |
| 446.4 | 446.4 | 1.00 | $A_{1g}$ | I | A |
| 1042.5 | 743.5 | 1.97 | $E_u$ | A | I |
| 1206.0 | 856.6 | 1.98 | $E_g$ | I | A |
| 1247.2 | 885.3 | 1.98 | $B_g$ | I | A |
| 1339.7 | 963.1 | 1.94 | $A_{2u}$ | A | I |

*4.3. FTIR spectroscopy*

Fig. 4 depicts the FTIR absorbance spectra ranging from 30 to 4000 cm$^{-1}$ of the Al-capped YH(D)O films, varying with the deposition pressure. Vibration bands at approximately 390, 515, 580, and 900 cm$^{-1}$ were observed for the YHO films deposited within the sputtering pressure range of 0.75–2.65 Pa. These bands are attributed to the vibrations of the YHO lattice; however, their considerable width suggests a highly disordered structure, which is also observed with ssNMR. Furthermore, the vibration bands are not affected by interference, as the broad peak at approximately 3650 cm$^{-1}$ corresponds to the first-order fringe. All these bands are also observable in the YDO films. The three lower-frequency bands at 390, 515, and 580 cm$^{-1}$ remain unshifted; however, the higher-frequency band at 900 cm$^{-1}$ experiences an isotopic shift to 745 cm$^{-1}$. These observations align with our calculations, indicating that bands below approximately 800 cm$^{-1}$ are primarily determined by Y and/or O oscillations and are minimally influenced by the substitution of H with D, whereas the higher bands are predominantly determined by H/D vibrations, thus experiencing isotopic shifts.

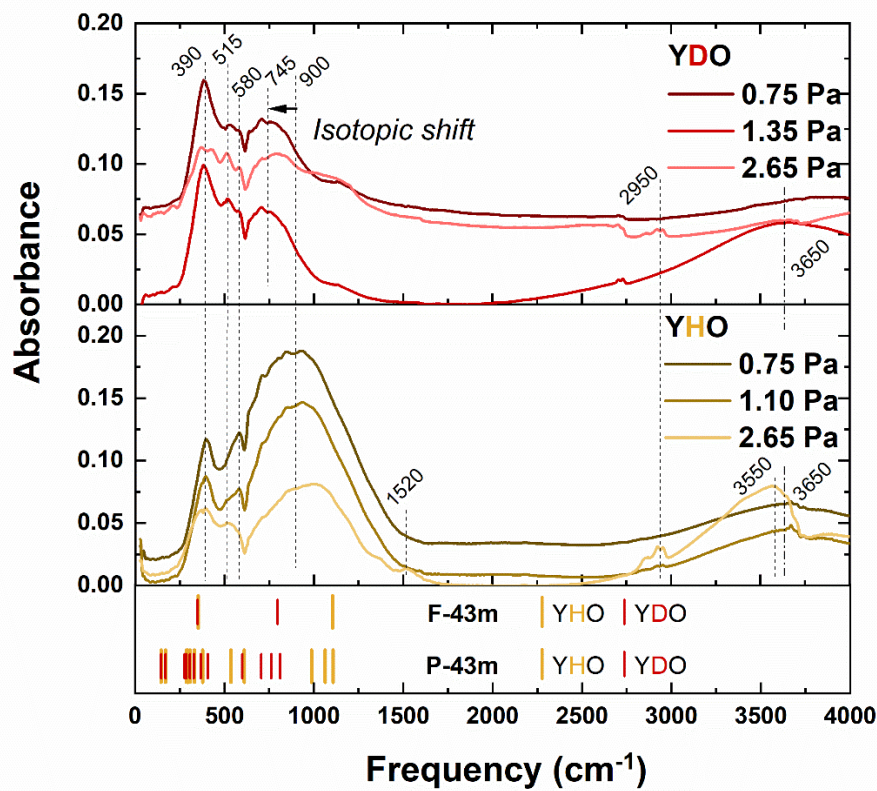

**Fig. 4.** Absorbance spectra (30–4000 cm$^{-1}$) measured in reflectance mode for Al-capped YH(D)O films deposited on Si substrates by reactive pulsed-DC magnetron sputtering (H$_2$(D$_2$)/Ar gas flow ratio of 1/2), varying with deposition pressure. The dashed lines indicate observed vibrational bands, with corresponding frequencies labelled above. The dashed-dotted line indicates the first-order interference fringe. The lower panel displays the calculated phonon mode frequencies (from Table 1) for two YH(D)O polymorphs, *F-43m* and *P-43m*.

The results of DFT calculations provide a qualitative framework for interpreting the measured FTIR spectra. Since the calculations were performed for ordered structures, their application is limited due to the broad absorbance bands observed experimentally, which indicate disorder within the anion sublattices. This disorder prevents precise identification of individual structures.

Previous studies have shown that a random arrangement of anions in tetrahedral sites is more energetically favourable [19], and experimental data confirm heterogeneous compositions [12] or even compositional gradients [15] in magnetron-sputtered films with post-oxidation.

In the most symmetric structure (*F*-43*m*), only two IR-active bands are predicted (Table 1), with one occurring at a significantly higher frequency (1106.6 cm$^{-1}$) than the experimentally observed broad band peaking at 900 cm$^{-1}$. The frequency of the 1106.6 cm$^{-1}$ mode is predominantly determined by hydrogen vibrations, as evidenced by the $\omega_H^2/\omega_D^2$ value of 1.93, which is close to the theoretical maximum of 2. For less symmetric structures, which contain more atoms in the unit cell, additional modes at the Brillouin zone centre appear at lower frequencies, some with $\omega_H^2/\omega_D^2$ values lower than 1.93, reflecting increased contributions from O and Y vibrations. The observed disorder in the anion sublattice can be interpreted as a structural deviation towards lower symmetry. As a result of this disorder, the experimentally observed band, which is not purely determined by hydrogen vibrations, appears at the lower frequency of 900 cm$^{-1}$, with an experimental $\omega_H^2/\omega_D^2$ value of approximately 1.45.

The potential formation of hydroxide (OH$^-$) or deuteroxide (OD$^-$) anions during growth has not been observed in films deposited below 1.35 Pa. Absorption bands at 1520, 2950, and 3550 cm$^{-1}$, attributed to OH bending, CH$_3$/CH$_2$ stretching, and OH stretching vibrations, respectively, are detected primarily in films deposited at higher pressures of 2.65 Pa, predominantly in the YHO case. The exact timing of the formation of these groups remains unclear. The bands at 1520 and 3550 cm$^{-1}$ (OH vibrations) have been previously observed [15] in non-capped films deposited at a pressure of 1.10 Pa but not in Al-capped films, suggesting that OH groups form during air exposure and that Al capping protects the films in this specific case. Conversely, the same study [15] demonstrated that films deposited at pressures higher than 1.10 Pa become partially transparent in the visible and near-infrared regions even during deposition, indicating oxidation and potential OH formation due to residual gases in the vacuum chamber.

Absorbance measurements are conducted in transmittance mode across the 30 to 7000 cm$^{-1}$ range for both the clear and darkened states of the non-capped YHO films deposited at 0.75 Pa (Fig. 5). The darkened state was achieved after 12 hours of sample illumination with UVA-violet light at 3.3 eV. Again, the broad peaks centred around 3350 cm$^{-1}$ and 6400 cm$^{-1}$ are due to interference effects, like those observed in Ref. [39] for YHO transmittance measurements in the mid-infrared region. However, a noticeable deviation from the smooth curve at approximately 3500 cm$^{-1}$ is attributed to OH stretching vibrations. The significant increase in absorbance (ΔA) following illumination is clearly visible in Fig. 5 and can be observed up to 2000 cm$^{-1}$ in the middle infrared region. At 7000 cm$^{-1}$, the absorbance increases from 0.07 (T=85%) to 0.21 (T=62%) after illumination. These measurements are carried out under both air and vacuum conditions for the photo-darkened film. Initially, the measurement is performed in air to minimise the time interval between measurement and cessation of illumination, as vacuuming requires time and YHO naturally undergoes bleaching. Consequently, the spectrum exhibits a noisy signal originating from uncompensated water in the spectrometer, rather than from the sample. The subsequent measurement in vacuum reveals a decrease in ΔA. Overall, phase transformation after illumination is not observed, as evidenced by the absence of significant changes in lattice-related vibration bands at 390, 540, and 850 cm$^{-1}$, which is consistent with findings from other studies [31]. The current measurement procedure does not allow for the assessment of potential subtle changes or changes with a small absorption cross-section that might occur if photo-darkening is induced by metallic cluster formation [40], subsequent local hydrogen displacement [41], and lattice parameter changes [42]. Indeed, a metallic cluster volume fraction of only 2% in the films would decrease transparency by over 30% [43]. More sophisticated in-operando FTIR measurements are necessary, albeit beyond the scope of this study.

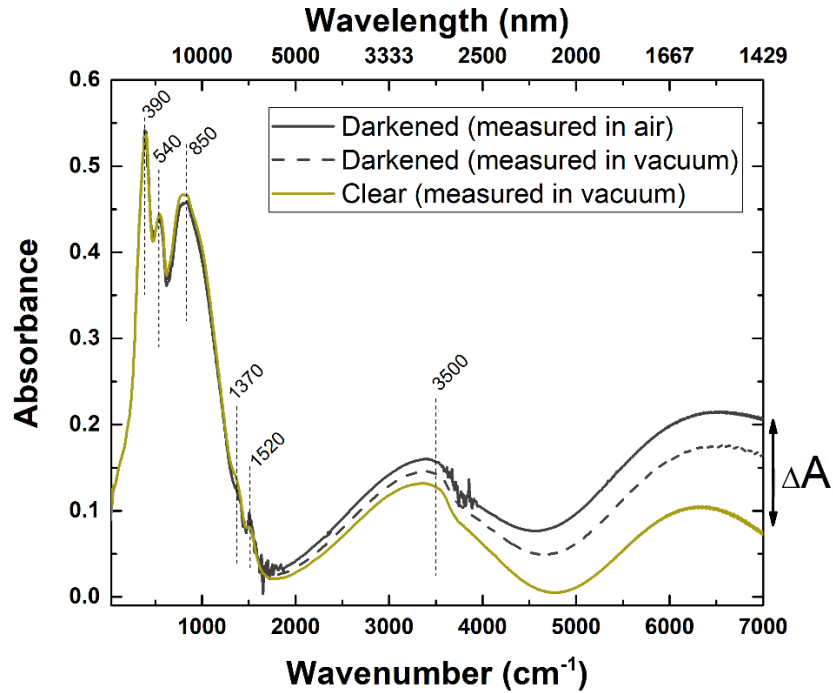

**Fig. 5.** Absorbance spectra (ranging from 30 to 7000 cm$^{-1}$) measured in the transmittance mode for the clear and darkened states of the non-capped YHO films deposited on Si at 0.75 Pa using reactive pulsed-DC magnetron sputtering (H$_2$(D$_2$)/Ar gas flow ratio of 1/2). Dashed lines represent observed bands, with corresponding frequencies indicated above.

The bands corresponding to OH groups at around 1370, 1520, and 3500 cm$^{-1}$ in Fig. 5 are more prominent in comparison to Al-capped films (Fig. 4), suggesting that the presence of OH groups is a result of exposure to air. However, following illumination, no increase in the intensity of these bands is observed, thereby disapproving the proposed mechanism discussed in Ref. [44], which links photo-darkening to the formation of OH groups. This finding also aligns with a DFT study in Ref. [45], which reports unstable OH species in YHO.

**Conclusions**

Both YHO and anion-labelled YDO films have been successfully deposited and characterised by FTIR spectroscopy in the far-infrared region for the first time, complemented by theoretical calculations of vibrational bands. The YH(D)O films are polycrystalline with a cubic crystal structure and small crystallite size in the range of 5–10 nm. The presence of a distribution of $^2$H anion environments, confirmed by ssNMR, leads to the extremely broad vibration bands in the FTIR spectra. It should be noted that powder-type samples could exhibit slightly different structure and composition due to the prolonged deposition process, which generally influences the structure and morphology and, in this case, also the extent of oxidation. Hydrogen contributes significantly to the broad vibration band around 850–900 cm$^{-1}$ in YHO films. The isotopic shift of this band is observed for the YDO films. The rest of the bands at lower frequencies are not influenced and are mainly determined by the Y and/or O movement. The above are confirmed both experimental FTIR measurements and theoretical calculations. Furthermore, our investigation of the photo-darkened films revealed a significant increase in absorbance following irradiation in the middle infrared region up to 2000 cm$^{-1}$. Despite these observed changes, no crystalline lattice transformation has been detected, as evidenced by the absence of significant alterations in lattice-related vibration bands. However, our measurement procedure did not allow for the assessment of potential subtle changes or changes with small absorption cross-section. Our findings not only contribute to the understanding of the vibrational properties of these photochromic films but also highlight the need for further investigations to elucidate the underlying mechanisms driving their photochromic behaviour.

**Acknowledgements**

Financial support was provided by Latvian Council of Science Project No. lzp-2022/1-0454. The Institute of Solid State Physics, University of Latvia, as a Center of Excellence, has received funding from the European Union's Horizon 2020 Framework Programme H2020-WIDESPREAD-01-2016-2017-TeamingPhase2 under grant agreement No. 739508, project CAMART$^2$.